\documentclass[acmlarge,nonacm]{acmart}
\usepackage[utf8]{inputenc}
\usepackage[T1]{fontenc}
\usepackage[font=itshape,indentfirst=false]{quoting}

\usepackage{amsfonts}
\usepackage{booktabs}
\usepackage{multirow}
\usepackage{icomma}
\usepackage{phdquery}
\usepackage{amsmath}
\usepackage{subcaption}
\usepackage{tabularx}
\usepackage{adjustbox}

\usepackage{phdmisc}

\usepackage{graphicx}
\graphicspath{{figures/}}

\begin{document}

\title{Fatigued Random Walks in Hypergraphs}
\subtitle{A Neuronal Analogy to Improve Retrieval Performance}

\author{José Devezas}
\email{jld@fe.up.pt}
\orcid{0000-0003-2780-2719}

\author{Sérgio Nunes}
\email{ssn@fe.up.pt}
\orcid{0000-0002-2693-988X}

\affiliation{%
  \institution{INESC TEC \& Faculty of Engineering, University of Porto}
  \streetaddress{Rua Dr. Roberto Frias, s/n}
  \postcode{4200-465}
  \city{Porto}
  \country{Portugal}}

\begin{abstract}
  Hypergraphs are data structures capable of capturing supra-dyadic relations. We can use them to model binary relations, but also to model groups of entities, as well as the intersections between these groups or the contained subgroups. In previous work, we explored the usage of hypergraphs as an indexing data structure, in particular one that was capable of seamlessly integrating text, entities and their relations to support entity-oriented search tasks. As more information is added to the hypergraph, however, it not only increases in size, but it also becomes denser, making the task of efficiently ranking nodes or hyperedges more complex. Random walks can effectively capture network structure, without compromising performance, or at least providing a tunable balance between efficiency and effectiveness, within a nondeterministic universe. For a higher effectiveness, a higher number of random walks is usually required, which often results in lower efficiency. Inspired by von Neumann and the neuron in the brain, we propose and study the usage of node and hyperedge fatigue as a way to temporarily constrict random walks during keyword-based ad hoc retrieval. We found that we were able to improve search time by a factor of $32$, but also worsen MAP by a factor of $8$. Moreover, by distinguishing between fatigue in nodes and hyperedges, we are able to find that, for hyperedge ranking tasks, we consistently obtained lower MAP scores when increasing fatigue for nodes. On the other hand, the overall impact of hyperedge fatigue was slightly positive, although it also slightly worsened efficiency.
\end{abstract}

\keywords{entity-oriented search, document ranking, hypergraph-based models, random walk score, neuronal fatigue analogy}

\maketitle

\section{Introduction}

A new era of computing is upon us\footnote{\url{https://thenextweb.com/contributors/2018/09/01/quantum-race-united-states-must-compete/}} and, while it might be a considerable challenge to deviate from well-established algorithms or data structures, such as the inverted index, it is also fundamental that we pursue alternative directions, in order to gain novel insights. This will promote creative ideas, enabling us to keep pushing the boundaries of information retrieval. Presently, search engines already aim to provide a more direct response to the information needs of users by, through query understanding, retrieving entity metadata, lists of entities or even direct answers to a question. This represents an improvement over the traditional list of ranked documents, obtained through a keyword-based query, making search more entity-oriented, in order to better serve the user's information needs~\cite{Bautin2007}.

There are three core tasks in entity-oriented search (EOS), that have been explored in TREC Entity track~\cite{Balog2010}, as well as INEX~XER track~\cite{Demartini2009}: ad hoc entity retrieval, related entity finding and entity list completion. Other tasks~\cite[Table 4.1]{Balog2018} include list search and similar entity search. Additionally, we also consider ad hoc document retrieval as an EOS task, but only when entities are factored into document ranking, which usually requires that both the query and the documents are semantically annotated. In this context, our goal is to find a representation and retrieval model capable of supporting the three core tasks along with ad hoc document retrieval when leveraging entities. The hypothesis is that, through a unified ranking function over a joint representation of text, entities and their relations, we will not only be able to provide a general implementation for entity-oriented search tasks, but also to improve retrieval effectiveness by maximizing the contribution of all available information.

Graphs and hypergraphs are particularly adequate data structures for the combination of text and knowledge representations. Hypergraphs, furthermore, have the advantage of being able to capture $n$-ary relations, higher-order dependencies and overlap, in a more natural manner. We have previously proposed a hypergraph-based representation and retrieval model for entity-oriented search tasks~\cite{Devezas2019}. Initial experiments with the hypergraph-of-entity had led to acceptable indexing times, but high search times. Depending on the collection, a query might take over $10$ minutes to run for random walks of length $\ell = 2$ and repeats $r = 10,000$~\cite[Table 6]{Devezas2019}. Although we could execute a query in approximately $10$ seconds for $\ell = 2$ and $r = 100$, at this point the algorithm wouldn't converge, thus outputting dissimilar rankings for repeated runs with the same configuration\footnote{Kendall's concordance coefficient $W \approx 0.84$ for 100 iterations with $\ell = 2$ and $r = 100$, as opposed to $W \approx 0.99$ with $\ell = 2$ and $r = 10,000$.}. This made the ranking function less useful, as it got closer to a random ranking of a random set of documents connected to the seed nodes (i.e., nodes representing the query in the hypergraph-of-entity). In order to make the hypergraph-of-entity useful in practice, we need to improve efficiency for large values of $r$, which correspond to points of high effectiveness and convergence\footnote{Notice that, when mentioning \emph{high effectiveness}, we are considering previous instances of the hypergraph-based model, instead of establishing a comparison to the state of the art, as we are still far from that target.}.

In this work, we adopt the idea of fatigue for random walks, introducing node and hyperedge fatigue as the number of cycles during which a given node or hyperedge is not available for visitation or traversal, as random steps are taken. Our initial hypothesis was that the addition of fatigue to the retrieval model would result in a significant overhead from maintaining and decrementing several variables of fatigue per cycle. On the other hand, regarding effectiveness, we didn't know what to expect --- we were led by the idea that, if this was a part of the neuronal process, then it should be relevant for effective cognition and thus might improve our ranking function. Contrary to our initial hypothesis and, as you will see in Section~\ref{sec:eval}, we found that the introduction of node fatigue significantly improved efficiency, however it decreased effectiveness, maintaining the tradeoff that we had evidenced before.

In Section~\ref{sec:ref-work}, we will cover reference work that inspired and supported our hypergraph-based representation and retrieval model. In Section~\ref{sec:rank}, we will describe the document ranking approach based on random walks, showing how fatigue is included in our model. In Section~\ref{sec:eval}, we will describe the test collection, as well as provide an assessment of efficiency and effectiveness, along with the measurement of the impact of node and hyperedge fatigue, and quantify the difference between a model without fatigue and a model with node fatigue. Finally, in Section~\ref{sec:concl}, we will close with final remarks and conclusions.

\section{Reference Work}
\label{sec:ref-work}

In this section, we begin by describing von Neumann's mention of fatigue, covering an existing application in computer science. We then deviate a bit to describe the way nondeterminism and random sampling is influencing computation, in order to increase efficiency with negligible impact for effectiveness. Finally, we delve into hypergraph-based models, covering text representation and retrieval modeling.

\subsection{Neuronal Fatigue in Computer Science}
\label{sec:ref-work:neuronal}

In preparation for Yale's Silliman Memorial Lectures\footnote{\url{https://en.wikipedia.org/wiki/Silliman_Memorial_Lectures}}, von Neumann highlighted the importance of jointly studying the computer and the brain~\cite{Neumann2012}, a work to be published posthumously for the first time in 1958. In fact, with his lecture, he provided enough common ground for crossover work between computer science and neuroscience, bringing the areas closer together. One of the ideas studied in von Neumann's lecture was the fact that a neuron will become fatigued, for a period of time, after having been stimulated.

\medskip

\begin{quoting}
  However, this is not the most significant way to define the reaction time of a neuron, when viewed as an active organ in a logical machine. The reason for this is that immediately after the stimulated pulse has become evident, the stimulated neuron has not yet reverted to its original, prestimulation condition. It is \emph{fatigued}, i.e.\ it could not immediately accept stimulation by another pulse and respond in the standard way. [\ldots] It should be noted that this recovery from fatigue is a gradual one [\ldots]''

  \quotingauthor{John von Neumann, The Computer and the Brain}
\end{quoting}

\medskip

\noindent Despite the impact neuroscience has had in computer science (e.g., neural networks), not many analogies with fatigue have been proposed. In fact, to our knowledge, only Xu and Yu~\cite{Xu2010} have used fatigue, in the context of neural networks, as a part of a revised version of backpropagation for spam filtering. In this work, we apply fatigue to nodes and hyperedges in a hypergraph, treating them as neurons that have been ``stimulated'' by a traversal during a random walk.

\subsection{Efficiency through Nondeterminism}

With the recent focus on quantum computing, it has become increasingly interesting to explore nondeterministic (but nonetheless converging) approaches, such as random walks. It is also curious to notice how some improvements in classical computing are being driven by quantum computing. An example is the work by Tang~\cite{Tang2018}, where he was able to propose a recommendation algorithm with a similar performance to the quantum state-of-the-art by Kerenidis and Prakash~\cite{Kerenidis2017}. In entity-oriented search (EOS), there are tasks analogous to recommendation, namely related entity finding (REF) and entity list completion (ELC). Our longterm goal is to propose a hypergraph-based representation and retrieval model for the generalization of EOS tasks, including ad~hoc document retrieval (explored in this work) and ad hoc entity retrieval. The strategy used by Tang~\cite{Tang2018} and Kerenidis and Prakash~\cite{Kerenidis2017} was based on a randomized sample from the user's preferences. Random walks in a hypergraph~\cite{Bellaachia2013} can be seen as a form of randomized sampling of the structure of the hypergraph. The longer the random walk or the higher the number of repeats, the better the ability to capture hypergraph structure. This means that we can easily decide on the trade-off between effectiveness and efficiency, simply by configuring the corresponding parameters. It also means that, assuming random walks do their job of correctly sampling structure, the representation model will then be the fundamental indicator of success of whatever task is implemented, which in our case is, as we have stated, the EOS task of ad hoc document retrieval through entity leveraging. Hypergraph structure acts as a constraint for random walking --- while random walks in the Euclidean space can essentially take a step in any direction, in hypergraphs they are restricted to taking steps within the structure of the hypergraph. The idea of fatigue that we explore here is simply an added restriction, similar to the one we introduce when moving from the Euclidean space to a hypergraph space.

\subsection{Hypergraph-Based Models}

Dekker and Birnbaum~\cite{HaentjensDekker2017} have proposed a hypergraph-based model for text, as an alternative to the XML tree paradigm. In their model --- Text As Graph (TAG) --- they generally included a \nodetype{document} node pointing to the first \nodetype{text} node. Each \nodetype{text} node, frequently for a single term, would then link to the following \nodetype{text} node, usually another term in the document, through a directed edge. Other, more complex structures would be captured using hyperedges like \edgetype{phrase}, \edgetype{line}, \edgetype{quatrain} or \edgetype{poem}. While TAG is a document-based representation, it is not hard to extend this model to a collection-based representation, such as the one we use here. Moreover, term or concept co-occurrence in phrases, lines or paragraphs, specially for a whole corpus, can act as relevance indicators, which is something that should be investigated in retrieval tasks. In 2012, Bendersky and Croft~\cite{Bendersky2012} proposed a query hypergraph to represent higher-order dependencies between terms in verbose queries. This differs from our work in two main aspects. First, we represent the collection of documents and associated knowledge as a hypergraph, mapping the query terms to nodes in the hypergraph, as opposed to building a hypergraph over query terms to act as a probabilistic graphical model. Secondly, we distinguish between terms and entities, which are, in our model, represented by two types of nodes, as opposed to representing an entity as a hyperedge over a set of term nodes --- removing \nodetype{entity} nodes is, however, something that we are considering doing in the future. The main advantage of a collection-based hypergraph is the ability to capture whichever information is available in the complete set of documents, harnessing the collective intelligence of different authors to cross-reference information as an inherent part of the ranking process.

\section{Node Ranking using Random Walks with Fatigue}
\label{sec:rank}

Random walks have been at the core of centrality metrics like PageRank or personalized PageRank~\cite{Page1999}. While PageRank is computed for the whole graph, personalized PageRank is computed for a localized area of the graph, based on a set of seed nodes. In both algorithms there is the probability that we follow a random outgoing edge. Given the complementary event that we don't, in PageRank we jump to a random node in the graph, but in personalized PageRank we always jump to one of the seed nodes instead, resulting in a behavior analogous to a random walk with restart~\cite{Tong2006}. The random walk score ranking function is similar to personalized PageRank, where our personalization is based on a keyword query and we use fixed length random walks, starting from each seed node, that only jumps back to its seed node after $\ell$ steps instead of doing it randomly. Each seed node represents an expansion of a query term to the entities it might refer to in the hypergraph (i.e., we follow all directed links between a query term and its neighboring entities). As such, departing from closer together seed nodes will reinforce the weight of the nodes/hyperedges they all cover, while departing from further apart seed nodes will lead to a middle-ground, reinforcing the weights of nodes/hyperedges in the intersecting borders instead.

The idea is to first reach an open interpretation of the query and, only then, as a part of the ranking process, close in on the actual sense of the query. By cross-referencing information based on all query terms, we attempt to diminish ambiguity, while at the same time performing ranking. For example, if we search for \query{Eiffel Tower}, we will expand to multiple entities mentioning \val{Eiffel} (at least \entity{Gustave Eiffel} and \entity{Eiffel Tower}) and multiple entities mentioning \val{Tower} (e.g., \entity{Eiffel Tower}, \entity{First National Bank Tower}, \entity{The Regal Tower}). It is through the combination of the neighborhoods of expanded-to entities that we can understand that the query is referring to \entity{Eiffel Tower}. While this is a straightforward example, with little ambiguity, more complex queries can only be segmented probabilistically based on existing knowledge~\cite{Devezas2016}. Take for instance \query{Gustave Eiffel tower construction}, where we could, at the very least, consider both \entity{Gustave Eiffel} and \entity{Eiffel tower} as valid (but overlapping) entities, despite only one of the options being meant by the user. On the other hand, the user might not provide enough information for disambiguation and query understanding, in which case the best option for the search engine is to consider the most probable segmentation and interpretation according to the corpus and/or knowledge base.

Using the seed nodes as an expanded representation of the query in the hypergraph, we then launch a random walk of length $\ell$ from each seed node, repeating this process $r$ times. Our expectation is that, given the appropriate hypergraph structure and restrictions (e.g., node/hyperedge weights, fatigue), we should be able to converge at a ranking of nodes/hyperedges based on the visitation frequency, for a sufficiently large $r$. We then collect the nodes and/or hyperedges that represent the target unit(s) of retrieval (documents and/or entities). We call this ranking function the random walk score, $RWS(Q, \mathcal{H}, \ell, r)$, for a given query $Q$ and hypergraph $\mathcal{H} = (V, E)$, where $V$ is the set of all vertices and $E$ is the set of all hyperedges, with $E_j$ either being a set of vertices, for undirected hyperedges, or a tuple with two sets of vertices, for directed hyperedges.

Previous experiments with $RWS$, namely in TREC 2018 Common Core track~\cite{Devezas2018}, have resulted in low or inconsistent effectiveness and worrisome efficiency, where queries frequently required several minutes to run, in order to converge to a stable ranking. The hypergraph-of-entity is a collection-based representation that links all available data, structured and unstructured, for a given corpus. This means that, whenever we query the hypergraph, we have access to a complete body of knowledge, but also that we will potentially be required to traverse a high number of paths before converging to a ranking. The question is then how to reduce the number of traversed paths without impacting effectiveness. In this section, we introduce and test fatigue in random walks in order to determine whether it can improve efficiency by acting as a controller for exploration of the untraveled paths. Fatigue as a restriction is not unlike the seed node introduced in personalized PageRank to focus on local exploration.

In order to propose a fatigued extension of random walk score, let us now consider von Neumann's description of fatigue (Section~\ref{sec:ref-work:neuronal}) and a \mbox{neuron--node} / \mbox{neuron--hyperedge} analogy. This tells us that, immediately after a node or \mbox{hyperedge} is traversed, it should enter a state of fatigue, and thus block random walk visitations for a given period of time $\Delta_{nf}$ (node fatigue) or $\Delta_{ef}$ (hyperedge fatigue). Time passes with each random step taken, meaning that, for every visited node $i \in V$ and hyperedge $j \in E$, we must initialize and maintain a hash table, respectively with $\scriptstyle\Delta_{nf}^i$ and $\scriptstyle\Delta_{ef}^j$ fatigue statuses that are decremented at each step. Each fatigued element is then excluded from the random sampling, when deciding which node or hyperedge to visit next at a random step. An element stops being fatigued when its fatigue status reaches zero, at that point being removed from the appropriate hash table. The addition of fatigue to the retrieval model results in a random walk score $RWS(Q, \mathcal{H}, \ell, r, \Delta_{nf}, \Delta_{ef})$, extended with the number of steps of fatigue for nodes and hyperedges, where $RWS(Q, \mathcal{H}, \ell, r) = RWS(Q, \mathcal{H}, \ell, r, \Delta_{nf} = 0, \Delta_{ef} = 0)$.

\section{Evaluation}
\label{sec:eval}

In this section, we present the INEX test collection, used to assess performance, and the analysis of retrieval efficiency and effectiveness for different representation and retrieval models. In particular, we compare two Lucene baselines, based on TF-IDF and BM25, with multiple configurations of RWS over two different hypergraph-based models, focusing on diversifying node and hyperedge fatigue, in order to understand their impact and behavior. Finally, we analyze the ranking correlations for Lucene BM25 and the best RWS, and for RWS with and without fatigue.

\subsection{Test Collection}

The INEX 2009 Wikipedia Collection is an XML collection of Wiki\-pe\-dia articles, which have been annotated with over $5,800$ entity classes from the YAGO~\cite{Suchanek2007} ontology (e.g., \type{person}, \type{movie}, \type{city}). It contains over 2.6 million articles and requires 50.7 GB of storage, when uncompressed. Based on this test collection, the INEX Ad Hoc track~\cite{Arvola2010} also provides 115 topics from the 2009 occurrence, with $50,725$ individual relevance judgments, and 107 topics from the 2010 occurrence, with $39,031$ individual relevance judgments. Each individual relevance judgment contains the query identifier, the document identifier, the number of relevant characters, the offset of the best entry point (usually the first relevant passage) and offset--length pairs for the relevant passages. For our experiments, we prepared a smaller subset of the INEX 2009 collection. We sampled 10 topics, uniformly at random, from the 2010 Ad Hoc track, and filtered the relevance judgments to keep only entries regarding the sampled topics. Finally, we discarded documents that were not mentioned in the relevance judgments, by removing them from each of the four archives in the collection. Although the subset that we prepared for assessment was only based on 10 topics, it contained $7,487$ documents along with $7,504$ individual relevance judgments.

\subsection{Representation and Retrieval Models}

Using Lucene, we indexed the text block of the smaller subset of INEX 2009 Wikipedia Collection, using TF-IDF and BM25 with $k_1 = 1.2$ and $b = 0.75$ as our baselines. We then indexed the text and knowledge blocks (i.e., terms, entities and their relations) using three versions of the hypergraph-of-entity.

The first version, the base model, included \nodetype{term} nodes from tokenizing the text into unigrams and \nodetype{entity} nodes from inter-document links (in this collection, a document represents a concept from Wikipedia, which we can use as a knowledge base). \nodetype{Term} and \nodetype{entity} nodes within a document were connected by an undirected \edgetype{document} hyperedge. Sets of \nodetype{term} nodes were connected to an \nodetype{entity} node by a directed \edgetype{contained\_in} hyperedge. Sets of \nodetype{entity} nodes were connected via an undirected \edgetype{related\_to} hyperedge, based on per-document entity co-occurrence. The goal in the future is to improve \edgetype{related\_to} hyperedges to take into account more details about entity relations, preferably establishing supra-dyadic relations, so that we ensure the hypergraph remains condensed.

The second version of the hypergraph-of-entity simply extended the base model with \edgetype{synonym} and \edgetype{context} hyperedges. Synonymy relations were based on WordNet~\cite{Miller1995} noun synsets, adding unexisting synonym \nodetype{term} nodes as required. Contextual similarity relations were based on a similarity network from word2vec~\cite{Mikolov2013}, built using 2-nearest neighbors for a cosine similarity above $0.5$.

The third version further extended the \edgetype{synonyms} and \edgetype{context} model with the node and hyperedge weights computed according to the functions proposed in the original work~\cite[Table 1]{Devezas2019}. For \nodetype{term} and \nodetype{entity} nodes, we used the sigmoid IDF; for \edgetype{document} hyperedges, we used a constant weight of 0.5; for \edgetype{related\_to} hyperedges, we used an average of the fraction of reachable other entities; for \edgetype{contained\_in},  we used the inverse size of the tail set (i.e., number of terms); for \edgetype{synonym}, we used the inverse number of synonyms it contained; and for \edgetype{context}, we used the average over all similarities between the original term and the contextual neighbor term, according to the cosine similarity between word embeddings.

For all tested hypergraph-of-entity models, we used RWS with length $\ell = 2$ and repeats $r = 1,000$. We then experimented with all combinations of node and hyperedge fatigue cycles within $\Delta_{nf} \in \{0, 10\}$ and $\Delta_{ef} \in \{0, 10\}$, resulting in four runs for each hypergraph-of-entity model, in addition to the two baseline runs. Our assessment was based on MAP, as well as P@10. We also measured indexing and search times to capture the impact that different models and parameter configurations had in efficiency.

\subsection{Performance Assessment}
\label{sec:perf}

Our assessment was based on the mean average precision (MAP), as well as the precision for the top 10 results (P@10). We also measured indexing and search times to capture the impact that different models and parameter configurations had in efficiency. Results are detailed in Table~\ref{tab:eval}.

\begin{table}[t]
  \centering
  \caption{Performance of random walk score, with different levels of fatigue, over the hypergraph-of-entity.}
  \label{tab:eval}

  \begin{tabular}{llccrr}
    \toprule
    \multicolumn{2}{l}{\multirow{2}{*}{\textbf{Version}}} & \multirow{2}{*}{\textbf{MAP}} & \multirow{2}{*}{\textbf{P@10}} & \multicolumn{2}{c}{\textbf{Search Time}}\\
      & & & & Avg./Query & Total\\
    \midrule
    \multicolumn{2}{l}{Lucene TF-IDF}       & 0.2160 & 0.2800 & 451ms       & 4s 510ms\\
    \multicolumn{2}{l}{Lucene BM25}         & 0.3412 & 0.4900 & 269ms       & 2s 688ms\\
    \midrule
    \multicolumn{6}{c}{\textbf{Hypergraph-of-Entity Base Model} -- $RWS(\ell = 2, r = 1,000)$}\\
    \midrule
    $\Delta_{nf}=0$   & $\Delta_{ef}=0$     & 0.1560 & 0.1800 & 1m 14s      & 12m 18s\\
    $\Delta_{nf}=0$   & $\Delta_{ef}=10$    & 0.1601 & 0.2300 & 1m 22s      & 13m 39s\\
    $\Delta_{nf}=10$  & $\Delta_{ef}=0$     & 0.0249 & 0.0900 & 839ms       & 8s 387ms\\
    $\Delta_{nf}=10$  & $\Delta_{ef}=10$    & 0.0246 & 0.1000 & 834ms       & 8s 338ms\\
    \midrule
    \multicolumn{6}{c}{\textbf{Hypergraph-of-Entity w/ Syns. \& Context} -- $RWS(\ell = 2, r = 1,000)$}\\
    \midrule
    $\Delta_{nf}=0$   & $\Delta_{ef}=0$     & 0.1594 & 0.2300 & 1m 05s      & 10m 48s\\
    $\Delta_{nf}=0$   & $\Delta_{ef}=10$    & 0.1540 & 0.2000 & 1m 05s      & 10m 46s\\
    $\Delta_{nf}=10$  & $\Delta_{ef}=0$ 	& 0.0236 & 0.0900 & 955ms       & 9s 553ms\\
    $\Delta_{nf}=10$  & $\Delta_{ef}=10$    & 0.0272 & 0.1100 & 924ms       & 9s 242ms\\
    \midrule
    \multicolumn{6}{c}{\textbf{Hypergraph-of-Entity w/ Syns., Context \& Weights} -- $RWS(\ell = 2, r = 1,000)$}\\
    \midrule
    $\Delta_{nf}=0$   & $\Delta_{ef}=0$     & 0.1636 & 0.2300 & 5m 26s      & 54m 15s\\
    $\Delta_{nf}=0$   & $\Delta_{ef}=10$    & 0.1615 & 0.1900 & 5m 31s      & 55m 13s\\
    $\Delta_{nf}=10$  & $\Delta_{ef}=0$ 	& 0.0195 & 0.0700 & 1s 011ms    & 10s 106ms\\
    $\Delta_{nf}=10$  & $\Delta_{ef}=10$    & 0.0250 & 0.1200 & 1s 049ms    & 10s 491ms\\
    \bottomrule
  \end{tabular}
\end{table}

\subsubsection{Efficiency}

Indexing with Lucene took $25s~889ms$, averaging $3.50ms$ per document. Indexing with the hypergraph-of-entity took $1m~9s~687ms$, averaging $9.42ms$ per document, for the base model. The times were similar for the extended models, taking $1m~4s~135ms$ and $1m~17s~193ms$, for an average of $8.67ms$ and $10.43ms$ per document, respectively for the non-weighted and weighted versions. Regarding search time, Lucene took on average, per query, $451ms$ for TF-IDF and $269ms$ for BM25. Hypergraph-of-entity was less efficient during search, taking on average, per query, between $1m~4s~620ms$ and $5m~31s~286ms$ for the base and extended models, when $\Delta_{nf} = 0$, that is, either without fatigue or with hyperedge fatigue only. On the other hand, for $\Delta_{nf} = 10$, average query time ranged between $834ms$ and $1s~49ms$, but it also resulted in a lower overall effectiveness.

\begin{figure}[t]
  \centering
  \includegraphics[width=\columnwidth]{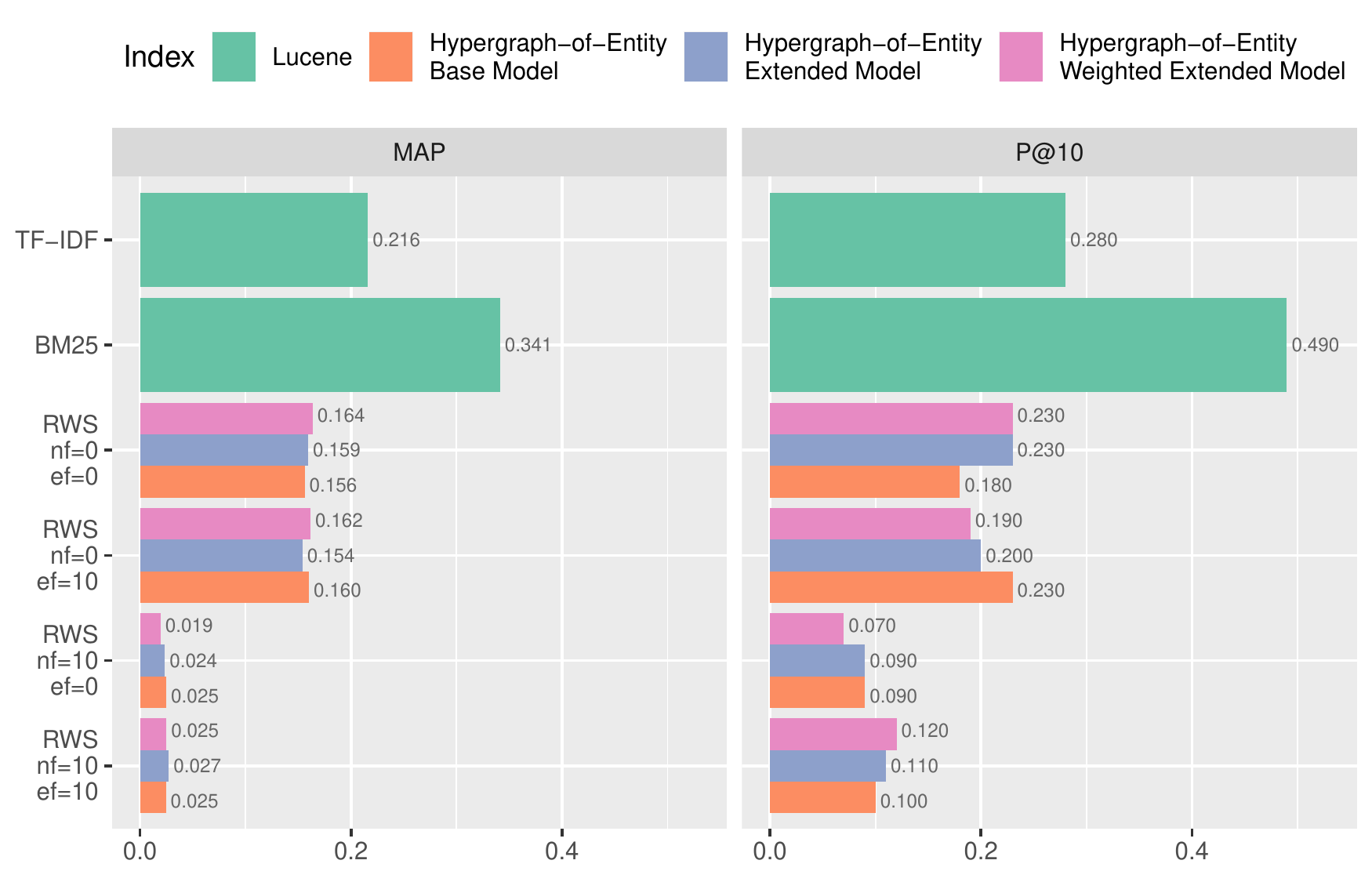}
  \caption{Evaluation metrics for Lucene baselines and hypergraph-of-entity models using different combinations of node and hyperedge fatigue for RWS.}
  \label{fig:eval}
\end{figure}

\subsubsection{Effectiveness}

We experimented with all four different combinations of node and hyperedge fatigue: no fatigue ($\Delta_{nf} = 0$, $\Delta_{ef} = 0$), node fatigue ($\Delta_{nf} = 10$, $\Delta_{ef} = 0$), hyperedge fatigue ($\Delta_{nf} = 0$, $\Delta_{ef} = 10$) and full fatigue ($\Delta_{nf} = 10$, $\Delta_{ef} = 10$). Figure~\ref{fig:eval} shows the MAP and P@10 scores for the hypergraph-of-entity base model, extended model and weighted extended model, when compared to the Lucene TF-IDF and BM25 baselines. A different filling color was used for each index (i.e., for each representation model). As we can see, none of the versions of the hypergraph-of-entity with RWS were able to outperform \mbox{TF-IDF} or BM25. Moreover, node fatigue had a negative effect in performance, while hyperedge fatigue had little to no effect on performance. We also found that the difference in MAP was not statistically significant when comparing \mbox{TF-IDF} and the best runs for RWS (base model with hyperedge fatigue and weighted extended model without fatigue). We obtained $p$-values of $0.3930$ and $0.4813$, respectively, for the Mann-Whitney U tests. On the other hand, when comparing the MAP for BM25 and the best run for RWS, we found that the difference was statistically significant, with a $p$-value of $0.004$ in both cases.

As it stands, the introduction of fatigue had little impact, except when considering the P@10 for the base model when using hyperedge fatigue. For that particular case, we were able to increase the performance of the base model without the need for synonyms, context or weights. Despite the small size of the sample, we were, at the very least, able to achieve a similar performance for \mbox{TF-IDF} and RWS, using a hypergraph-based model, instead of an inverted index, and a nondeterministic random walk based approach. The model we propose has the potential to, through its joint representation of text, entities and their relations, unlock novel ranking strategies that take into account all available leads, be it those locked within unstructured text or those explicitly provided through structured knowledge. Beyond document ranking, hypergraph-of-entity can easily support entity ranking, related entity finding and entity list completion.

\subsection{Rank Correlation Analysis}

We used Spearman's rank correlation coefficient $\rho$ to compare vectors of positions without ties. Any missing positions were added to either vector, using the lexicographical order for tied documents, based on their ID. Missing documents were assigned synthetic incremental positions, after the last retrieved document, in order to complete the rankings and make them comparable. Given the nondeterministic (but overall converging) character of RWS, we computed and averaged 100 $\rho$ values for the same parameter configuration, in order to obtain a robust insight.

\begin{table}[t]
  \caption{Spearman's rank correlation coefficient $\rho$ and Jaccard index $J$, averaged over 100 repeated retrieval events for the same topic. \small\normalfont $\left<\rho_1\right>$ and $\left<J_1\right>$ compare TF-IDF and the best RWS with fatigue, while $\left<\rho_2\right>$ and $\left<J_2\right>$ compare RWS with fatigue ($\Delta_{ef} = 10$) and without fatigue. The mean $\mu$ and standard deviation $\sigma$ are shown at the bottom of the table.}
  \label{tab:spearman}

  \begin{tabularx}{\linewidth}{lRRRR}
    \toprule
    \textbf{Topic} & $\left<\rho_1\right>$ & $\left<\rho_2\right>$ & $\left<J_1\right>$ & $\left<J_2\right>$\\
    \midrule
    2010003	& -0.8400 &  0.9639 & 0.0000 & 0.8818\\
    2010014	& -0.7707 &  0.9961 & 0.0000 & 1.0000\\
    2010023	& -0.6867 & -0.1238 & 0.0000 & 0.2323\\
    2010032	& -0.7256 & -0.4649 & 0.0147 & 0.1650\\
    2010038	& -0.7539 &  0.8086 & 0.0316 & 0.7475\\
    2010040	& -0.7740 &  0.9788 & 0.0000 & 0.8806\\
    2010049	& -0.7455 &  0.9460 & 0.0000 & 0.8324\\
    2010057	& -0.6500 &  0.9242 & 0.0526 & 0.8011\\
    2010079	& -0.7295 &  0.9730 & 0.0000 & 0.9382\\
    2010096	& -0.6864 &  0.8871 & 0.0000 & 0.7487\\
    \bottomrule
    \multicolumn{1}{r}{$\mu$}     & -0.7362 & 0.6889 & 0.0099 & 0.7228\\
    \multicolumn{1}{r}{$\sigma$}  & 0.0541  & 0.5272 & 0.0183 & 0.2876\\
  \end{tabularx}
\end{table}

We ran two experiments. The first enabled us to further understand the differences between the baselines and our models. We compared the rankings provided by Lucene TF-IDF, achieving a P@10 of $0.2800$, with the rankings provided by RWS, for $\ell = 2$, $r = 1,000$, $\Delta_{nf} = 0$ and $\Delta_{ef} = 10$, achieving a P@10 of $0.2300$ for the base model (the best with fatigue). In the second experiment, we compared two versions of RWS, with and without fatigue. Particularly, we focused on hyperedge fatigue, since it resulted in a similar performance to the version without fatigue. The second experiment was run over the extended model.

Table~\ref{tab:spearman} shows the average $\rho$ values, $\left<\rho_1\right>$ and $\left<\rho_2\right>$, for each experiment, per topic. It also shows the respective average Jaccard indexes, $\left<J_1\right>$ and $\left<J_2\right>$, as a complement to correlation analysis. At the end of the table, mean ($\mu$) and standard deviation ($\sigma$) values are shown to summarize global behavior. As we can see, when comparing TF-IDF and RWS, we obtained values for $\rho_1$ that consistently approximate $-1$, with a mean of $-0.7362 \pm 0.0541$, an indication that TF-IDF and RWS are anticorrelated. If we look at $\left<J_1\right>$, we find extremely low similarity values between the document sets returned by TF-IDF and RWS, with this value ranging around $0.0099 \pm 0.0183$. This explains the negative correlation and, interestingly, shows that RWS can still achieve good retrieval effectiveness while returning an almost completely different set of documents than TF-IDF. Regarding the comparison of RWS with and without fatigue, we were unable to find a recurrent pattern, like we did for in the first experiment. We found both positive and negative correlations, with lower Jaccard index values associated with negative correlations. It is, however, clear that the usage of fatigue in RWS results in a different ranking than the standard RWS without fatigue. The document set similarity between the two configurations, which is of $0.7228 \pm 0.2876$, is also higher than the first experiment, which is consistent with the fact that we are testing a different version of the same ranking function. The introduction of hyperedge fatigue particularly affected topics $2010023$ and $2010032$ for \query{retirement age} and \query{japanese ballerina}, respectively. They both achieved a similarly low P@10 ($0.0$ and $0.2$, versus $0.1$ and $0.3$, respectively, when comparing RWS with and without fatigue for the pair of topics); the behavior was similar for P@1000. Together with a low Jaccard index ($0$ for topic $2010023$ and $0.0147$ for topic $2010032$) this indicates that both approaches were able to retrieve different sets of relevant documents.

\section{Conclusion}
\label{sec:concl}

Inspired by von Neumann's last lecture and his motivation for crossover work between computer science and neuroscience, we have proposed the application of fatigue to random walks in hypergraphs. After stimulated, a neuron in the brain enters a state of fatigue that lasts a given period of time. We applied this analogy to nodes and hyperedges where, during a random walk, a node and/or hyperedge would be fatigued for a given number of cycles before a random walker could traverse it again. The random walk score (RWS) that we explore here is a ranking function used for entity-oriented search tasks, such as ad hoc document retrieval through the leveraging of entities. In order to rank documents, RWS first mapped the query to a set of seed nodes that best represented it in the hypergraph and launched multiple random walkers until a ranking was obtained through \edgetype{document} hyperedge visitation frequency.

We found that fatigue was able to significantly improve retrieval efficiency, at the cost of effectiveness, particularly when compared to the RWS version without fatigue. In the best case scenario, for the base model, search time improved from $1m~14s$ to $834ms$ but MAP decreased from $0.1560$ to $0.0246$. Behavior was similar for the extended model, but more extreme for the weighted extended model. For the weighted extended model, search time improved from $5m~26s$ to $1s~011ms$ but MAP decreased from $0.1636$ to $0.0195$. While we were unable to surpass the baselines, we were able to introduce fatigue and achieve a similar performance to the random walk score without fatigue. In particular, we were able to do so when introducing fatigue in hyperedges, which were the object of ranking, given they represented documents. In the future, it would be interesting to experiment with entity ranking and assess whether the impact of node fatigue is more positive than hyperedge fatigue, in a scenario where nodes are the object of ranking, given they represent entities.

Finally, through correlation analysis, we further investigated the similarities between the rankings obtained from different models. We specifically compared the TF-IDF baseline with the best RWS with fatigue. We found that RWS was able to effectively retrieve documents, reaching a comparable performance to the baselines, but returning a different set of relevant documents, in a nearly anticorrelated manner. Using a similar strategy, we also compared two configurations of RWS, with and without fatigue, also finding that they returned different sets of relevant documents, while showing an overall positive correlation, except for two topics, where only a few relevant results could be retrieved.

\begin{acks}
  José Devezas was supported by research grant PD/BD/128160/2016, provided by the Portuguese national funding agency for science, research and technology, Fundação para a Ciência e a Tecnologia (FCT), within the scope of Operational Program Human Capital (POCH), supported by the European Social Fund and by national funds from MCTES.
\end{acks}

\bibliographystyle{ACM-Reference-Format}
\bibliography{hgoe-fatigue.bib}

\end{document}